\documentclass{article}

\usepackage{euscript}

\usepackage{latexsym}
\usepackage{amsmath, amssymb,graphics}
\usepackage[all]{xy}

\usepackage{fullpage}

\newtheorem{theorem}{Theorem}

\newtheorem{lemma}[theorem]{Lemma}

\newtheorem{proposition}[theorem]{Proposition}
\newtheorem{conjecture}{Conjecture}

\renewcommand{\(}{\begin{equation*}}
\renewcommand{\)}{\end{equation*}}
\newcommand{\bea}{\begin{eqnarray*}}
\newcommand{\eea}{\end{eqnarray*}}
\newcommand{\R}{{\mathbb R}}
\newcommand{\C}{{\mathbb C}}
\newcommand{\Z}{{\mathbb Z}}

\renewcommand{\a}{\alpha}

\newcommand{\cC}{\ensuremath{\mathcal C}}
\newcommand{\cE}{\ensuremath{\mathcal E}}

\newcommand{\cG}{\ensuremath{\mathcal G}}
\newcommand{\cH}{\ensuremath{\mathcal H}}

\newcommand{\bo}{\raise-1mm\hbox{\Large$\Box$}}              
\def\i{\ensuremath{\dot\imath}}

\def\i{\ensuremath{\dot\imath}}

\newcommand{\beq}{\begin{equation}}
\newcommand{\eeq}{\end{equation}}

\numberwithin{equation}{section}

\renewcommand{\(}{\begin{equation}}
\renewcommand{\)}{\end{equation}}

\def\R{{\mathbb R}}
\def\Z{{\mathbb Z}}

\def\C{{\mathbb C}}

\def\1{{\bf 1}}

\def\<{\langle}
\def\>{\rangle}

\def\a{\alpha}

\numberwithin{equation}{section}

\renewcommand{\(}{\begin{equation}}
\renewcommand{\)}{\end{equation}}

\begin{document}

\begin{titlepage}


\vspace{2em}
\def\thefootnote{\fnsymbol{footnote}}

\begin{center}
{\Large\bf 
Duality and cohomology  in M-theory with boundary}
\end{center}
\vspace{1em}

\begin{center}
\large Hisham Sati 
\footnote{e-mail: {\tt
hsati@math.umd.edu}}
\end{center}

\begin{center}
Department of Mathematics\\
University of Maryland\\
College Park, MD 20742 
\end{center}

\vspace{0em}
\begin{abstract}
\noindent
We consider geometric and analytical aspects of M-theory on a manifold with 
boundary $Y^{11}$. 
The partition function of the C-field requires summing over harmonic forms.
When $Y^{11}$ is closed Hodge theory gives a unique harmonic form in each
de Rham  cohomology
class, while in the presence of a boundary the Hodge-Morrey-Friedrichs decomposition 
should be used. This leads us to study the boundary conditions for the
C-field. The dynamics and the presence of the dual to the C-field gives 
rise to a mixing of boundary conditions
with one being Dirichlet and the other being Neumann. We describe the mixing 
between the corresponding absolute and relative
cohomology classes 
via Poincar\'e duality angles, which we also illustrate for the M5-brane as a tubular 
neighborhood. Several global aspects are then considered. We provide a systematic
study of the extension of the $E_8$ bundle and characterize obstructions. Considering 
$Y^{11}$ as a fiber bundle, we describe how the phase looks like on the base, hence
providing dimensional reduction in the boundary case via the adiabatic limit of the eta invariant.
The general use of the index theorem leads to a new effect given by a gravitational Chern-Simons term $CS_{11}$ 
on $Y^{11}$ whose restriction to the boundary would be a generalized WZW model. 
This suggests that holographic models of M-theory can be viewed as 
a sector within this index-theoretic approach.

\end{abstract}

\end{titlepage}

\tableofcontents

\section{Introduction}

M-theory has proven to be a rich theory both in terms of modeling 
physical phenomena and in terms of mathematical structures. 
Physical and mathematical insight could be gained by studying various 
aspects of this theory.  
In this paper we study geometric and analytical 
aspects of M-theory on a manifold 
with a boundary, building on \cite{DFM}, 
mainly by emphasizing importance of boundary conditions 
and their effect on the corresponding bundles, fields, actions and partition
functions. The main ingredient we use for the kinetic term is 
harmonic forms. In the presence of a boundary, the Hodge decomposition theorem
has to be modified 
and new effects appear, depending on boundary conditions. 
For the phase we use index theory on manifolds with a boundary, departing
from \cite{DFM} and \cite{FM} by the use of the adiabatic limit of the eta-invariants.

\vspace{3mm}
We take M-theory on an 11-dimensional Spin manifold with 
boundary $Y^{11}$ equipped with a Riemannian  metric $g_Y$. 
The main fields we consider are the C-field $C_3$ with its field strength $G_4$
as well the dual field $G_7$,
the 11-dimensional Hodge dual to $G_4$ at the level of differential forms. 
The $C$-field has a classical harmonic part (see e.g. \cite{Gauss}), 
which is characterized in 
\cite{tcu} in the extension to the Spin bundle. 
The Bianchi identity and equation of motion for the $C$-field
in M-theory, which follow from those of eleven-dimensional supergravity \cite{CJS},
are 
 \(
 dG_4 =0\;, \qquad
\frac{1}{\ell_p^3} d*G_4 = \frac{1}{2}G_4 \wedge G_4 - I_8,
 \label{EOM}
 \)
where $I_8$ is the one-loop polynomial, 
$*$ 
is the Hodge duality operation in eleven dimensions,
and $\ell_p$ is the scale in the theory called the Planck `constant'.  
A formulation in terms of $G_4$ and $G_7$ is given in \cite{CJLP}.
The presence of $d \ast G_4=dG_7$ suggests looking also at a degree eight field
$G_8$ (this is called $\Theta$ in \cite{DFM}). The two fields $G_4$ and $G_8$
can be treated in a unified way \cite{S1} \cite{S2} \cite{S3} \cite{tcu}.

\paragraph{Harmonic C-field.} 
The classical (or low energy) limit 
is obtained by taking $\ell_p \to 0$ and is dominated by the metric-dependent 
terms. 
In this long distance approximation of M-theory
 one keeps only the harmonic modes of the 
C-field \cite{Gauss} \cite{tcu}. 
Let
$
\Delta_g^3 : \left(\Omega^3(Y^{11}), g \right)
\longrightarrow
\left(\Omega^3(Y^{11}), g \right)
$
be the Hodge Laplacian on 3-forms on the base $Y^{11}$ with respect
to the metric $g$ given by $\Delta_g^3= d~d^{{}^*}+d^{{}^*}d$, where
$d^{{}^*}$ is the adjoint operator to the de Rham differential operator 
$d$. 
Assuming $[G_4]=0$ in $H^4(Y^{11};\R)$ so that
$G_4=dC_3$ then in the Lorentz gauge, $d^{{}^*}C_3=0$, we have
\cite{tcu} $\Delta_g^3 C_3= * j_e$, where $j_e$ is the electric current associated
with the membrane given by
$
j_e= \ell_p^3 \left(\frac{1}{2} G_4 \wedge G_4 - I_8 \right)
$.
Thus, $C_3$ is harmonic if $\ell_p \to 0$ and/or there are no membranes.
The space of harmonic 3-forms on $Y^{11}$ is 
$
{\cH}_g^3 (Y^{11}) : = {\rm ker} \Delta_g^3 \subset \Omega^3(Y^{11})
$.
Harmonic forms are very important in compactification, where 
the fields are expanded in a harmonic basis. 
For instance, if $\a^i$ is basis for the space
$\cH^3(Y^{11})$ of harmonic 3-forms on $Y^{11}$ then the C-field can be expanded
as $C_3= \sum_i C_3^i \a^i$.
There are natural choices for internal manifolds for compactifications with fluxes
leading to supersymmetric theories in lower dimensions (see \cite{Du} and references therein).
A seven-dimenisonal manifold $M$ with a 3-form $\varphi$ 
is a $G_2$ manifold if $d \varphi= d^* \varphi=0$, that is if 
$\varphi$ is harmonic. 
An eight-manifold with a self-dual four-form $\phi=\ast \phi$ 
is called a torsion-free  Spin(7) manifold if $d\phi=0$.

\paragraph{The C-field in the presence of a boundary.}
When $Y^{11}$ has a boundary we no longer assume that there is 
a bounding twelve-manifold $Z^{12}$. 
The topological sectors of the C-field are labelled by 
extensions $\tilde{a}$ of the degree four
characteristic class of the C-field on $M^{10}=\partial Y^{11}$.
In addition to summing over torsion, there will be an integral over a certain 
space of harmonic fields. 
Consider the inclusion $i: M^{10} \hookrightarrow Y^{11}$, which induces 
the pullback on cohomology $i^*: H^4(Y^{11};\Z) \to H^4(M^{10};\Z)$. In \cite{DFM}
the sum over the topological sectors in the wavefunction 
is restricted to ${\rm ker}i^* \subset H^4(Y^{11};\Z)$, which is equivalent to a sum over
$H^4(Y^{11}, M^{10};\Z)/\delta H^3(M^{10};\Z)$, where $\delta$ is the connecting 
homomorphism, and the integration in the path integral would be
 over the compact space of harmonic forms
$
\cH^3(Y^{11}, M^{10})/\cH^3(Y^{11}, M^{10})_\Z$,
where $\cH^3(Y^{11},M^{10}):= {\rm ker} i^*$ restricted to 
$\cH^3(Y^{11})$. 
It is desirable to further characterize these, which is one of the goals of this paper. 
We formulate a boundary value problem which is solvable from general considerations
in section \ref{sec bvp}. We work with $C_3$ as well as its field strength
so that both degree three and four cohomology are relevant.

\paragraph{Boundary conditions and duality.}
In the absence of the field dual to the C-field, the boundary conditions 
can be taken in a straightforward way. However, when this field is 
introduced, an interplay between Hodge duality and dynamics of the 
fields makes such obvious choices not possible. 
In particular,  if the C-field is taken to satisfy the Dirichlet boundary condition
then its dual must satisfy the Neumann boundary condition, and vice versa.
Thus in this paper we provide a systematic study of these matters. 
Naturally, then one might ask what replaces the duality in $Y^{11}$ 
when one restricts to the boundary. We study analogs of the Hilbert 
transform introduced in \cite{BS} which effectively provides 
a description for such a duality and exchanges Dirichlet and Neumann 
forms. 
In addition, we will consider generalization to include the dual fields
in section \ref{sec hod}.

\paragraph{Cohomology in the presence of boundary.} 
An arbitrary de Rham cohomology class of an oriented compact Riemannian 
manifold can be represented by a unique harmonic form, i.e. the natural 
map $\cH^k(M) \to H^k_{dR}(M)$ is an isomorphism. This means that 
every cohomology class contains exactly one harmonic form. 
When $Y^{11}$ is closed then, from the Hodge decomposition theorem, the fourth
cohomology group with real coefficients $H^4(Y^{11};\R)$
is isomorphic to the space of closed and coclosed differential 4-forms on $Y^{11}$.
Thus, the space of these harmonic forms provides a concrete realization of the 
cohomology group $H^4(Y^{11};\R)$ inside the space $\Omega^4(Y^{11})$ of 
all 4-forms on $Y^{11}$. 
The Laplacian $\Delta$
on $p$-forms on a closed  $Y^{11}$ is self-adjoint. However, in the presence of a boundary this 
is no longer the case, and in fact $\Delta$ is surjective \cite{CDGM}. 
In this case we use (in section \ref{sec bvp})
the Hodge-Morrey-Friedrich (HMF) decomposition theorem
\cite{Mo} \cite{Fri} which gives us a concrete realization of the absolute 
cohomology 
$H^4(Y^{11};\R)$ and the relative 
cohomology $H^4(Y^{11}, \partial Y^{11};\R)$ inside the 
space of all harmonic 4-forms on $Y^{11}$. 
The two spaces, surprisingly, intersect only at zero (see \cite{Sh})
$
{\rm Harm}^4(Y^{11}, \partial Y^{11};\R) \cap {\rm Harm}^4(Y^{11};\R)=\{0\}$.
In addition, the boundary subspace of each is orthogonal to all of the 
other. 
Each of $H^4(Y^{11};\R)$ and $H^4(Y^{11}, \partial Y^{11};\R)$
has a portion consisting of those cohomology classes coming from the 
boundary $\partial Y^{11}$ and another portion of those coming from the
interior part of $Y^{11}$. 
The principal angles between the interior subspaces of the concrete 
realizations of $H^4(Y^{11};\R)$ and $H^4(Y^{11}, \partial Y^{11};\R)$
are called {\it Poincar\'e duality angles} and are characterized, via a refinement of the 
Hodge-Morrey-Friedrich decomposition,  
in \cite{DG} and also \cite{Sh}.
Poincar\'e duality angles measure how near
 a manifold with boundary is to being closed.
We will be interested in orthogonal decomposition and not just 
direct sum decomposition (kinetic terms etc.).
Similar discussion is provided for other fields, namely $C_3$, $G_7$ and $G_8$.
We highlight new effects on the fields due to such phenomena. 
All of this is discussed in section \ref{sec PD}.
Considering the M5-brane as a tubular neighborhood in the ambient spacetime
we illustrate, in section \ref{sec m5},
 the dependence of kinetic terms on distance scales.

\paragraph{$E_8$ gauge theory for $\partial Y^{11}\neq \emptyset$.}
The phase of the (non-gravitational) partition function
can be studied using $E_8$ gauge theory \cite{DMW}.
For each characteristic class $a\in H^4(Y^{11};\Z)$ of an $E_8$ bundle over $Y^{11}$
there is a harmonic 
four-form $G^a_4$ of the appropriate topological class. The kinetic 
energy 
$|G_4^a|^2=\int_{Y^{11}} G_4^a \wedge \ast G_4^a$ vanishes if and 
only if $G_4^a$ is torsion. 
The partition function involves evaluating the sum 
$
\sum_{a \in H^4(Y^{11};\Z)} (-1)^{f(a)}
\exp \left( -|G_4^a|^2 \right)$,
where $f(a)$ is a quadratic refinement of a bilinear form related to $a$ \cite{DMW}. 
In dealing with boundaries one has to impose boundary conditions on the 
C-field within the $E_8$ model. In \cite{DFM} the conditions $i^*(C)=0$ is chosen, where
$\check C=(A, C)$, $A$ is a connection on an $E_8$ bundle. 
This restricts to the C-fields ${\rm E}_P(Y^{11},M^{10}):=\{ (A,C) \in {\rm E}_P(Y^{11})~|~
i^* (C)=0 \}$. 
The boundary condition breaks the topological gauge symmetry $\cG$, which 
is breaking of groupoid rather than of structure groups. 
This implies the following for the kinetic terms of the $E_8$ gauge field strength $F$ 
\cite{DFM}. 
While the term $\int_{Y^{11}} {\rm tr} F \wedge \ast F$ is not gauge invariant, and hence has no 
physical degrees of freedom in the interior,  the corresponding term 
$\int_{M^{10}} {\rm tr} F \wedge \ast F$ on the boundary is gauge invariant and hence
defines dynamical $E_8$ gauge fields there. 
Some supersymmetric aspects of this are discussed in \cite{ES}.
The generalization of the boundary condition on the C-field  
leads to conditions for extension  of the $E_8$ bundle, which we also characterize
in section \ref{sec e8}. 
This generalizes the abelian case in four dimensions \cite{Zu} to the $E_8$ case
in eleven dimensions (but the discussion holds in more dimensions).

\paragraph{The effective action and partition function.}
The exponentiated action in the closed case is \cite{DMW}
\(
\exp \left[ 
-2\pi \int_{Y^{11}} \frac{1}{\ell_p^9}{\rm vol}(g_Y)R(g_Y) + \frac{1}{2\ell_p^3}
G_4 \wedge \ast G_4 
\right]  \Phi_{RS} \cdot \Phi (C_3)\;,
\label{exp ac}
\)
where 
$\Phi_{RS}$ is the 
Rarita-Schwinger contribution from the Dirac form $\overline{\psi} D \psi$,
 and $\Phi (C_3)$
is the phase built out of the topological parts of the action, namely the 
Chern-Simons term and the one-loop term. This phase is studied 
extensively in 
\cite{DMW} for the case of no boundary and in \cite{DFM} when 
$Y^{11}$ has a nonempty boundary.  
The phase also leads to interesting topological structures 
\cite{SSS2} \cite{tcu} \cite{DMW-sig}.
In this paper we concentrate both on the phase and 
on the terms involving $\ell_p$, and in 
particular the kinetic term for the C-field. 
We can see that when $\ell_p$
is small,
 corresponding to a semiclassical approximation, 
the kinetic term for the C-field will dominate 
the exponential in \eqref{exp ac} while the contribution from the 
Einstein-Hilbert term ($R(g_Y)$ is the scalar curvature) 
will be very small in comparison. 
This can also be seen from the measure. 
The measure for the C-field path integral is \cite{DFM}
$
\mu (C_3) \cdot {\rm Pfaff}(D_{RS}) \cdot \Phi(C_3) \exp [{-\frac{1}{\ell_p^3} \int_{Y^{11}} G_4 \wedge \ast G_4}]$,
where $\mu (C_3)$ is the standard formal measure for 3-form gauge potentials 
defined by the metric $g_Y$ on $Y^{11}$ together with the Faddeev-Popov 
procedure applied to small C-field gauge transformations. 
As in \cite{DFM} we also consider a fixed metric on $Y^{11}$.
In addition to the terms in \eqref{exp ac} there are terms corresponding to 
four-fermion interactions as well as interactions of fermions with 
the C-field. Aspects of the latter is considered in \cite{ES} and more fully in
\cite{LM}.

\paragraph{The phase via the adiabatic limit in the boundary case.}
The boundary conditions for the Dirac operator and the corresponding 
Pfaffians on $Y^{11}$ with boundary 
are discussed in detail in \cite{FM}. M-theory is shown to be well-defined 
topologically on an arbitrary number of spatial components. 
The partition function is constructed in the presence of boundaries
using exponentiated eta-invariants. 
What we do (in section \ref{sec eta}) 
is provide an interpretation using the adiabatic  limit as 
 in the case of no boundaries \cite{MS} \cite{S-gerbe} \cite{DMW-sig}.
Take $Y^{11}$ to be the total space of a bundle with fiber 
$N^n$ and base space$X^{11-n}$. 
Assuming $\partial Y^{11}\neq \emptyset$, there are two cases to consider.
First, that the base has a boundary, in which case we make use of the 
results of Dai \cite{D}. Second, that instead the fiber has a boundary, for which we realize 
the constructions of Bismut-Cheeger \cite{BC1} \cite{BC2} and of Melrose-Piazza
\cite{MP}. The adiabatic limit, via the point of view advocated in 
\cite{MS} \cite{S-gerbe} \cite{DMW-sig}, amounts to 
a dimensional reduction which keeps track of the geometry and analysis involved. Thus, the final 
expressions will be ones on the base $X^{11-n}$.

\paragraph{Geometric corrections to the index and the gravitational Chern-Simons term.}
Local boundary conditions (Dirichlet) can be imposed for the de Rham complex. 
However, for the Spin and signature complexes, this can no longer be done
due to topological obstructions to finding local boundary conditions. Instead, one 
uses the spectral boundary condition determined by the spectrum of operators on 
the boundary \cite{APS1}. 
Furthermore, in the case of differential forms with duality, as is the case with the C-field and its dual
described by the signature complex \cite{DMW-sig}, 
we can no longer impose Dirichlet boundary conditions on both the C-field and its dual. 
So there will be an exclusion principle. 
Furthermore, the proof in \cite{APS1} of the index theorem with boundary 
assumes that the Riemannian manifold has a product metric near the boundary. 
For general manifolds, there is a correction form given by 
a gravitational Chern-Simons term.
In \cite{Ho} Horava proposed a holographic 
nonperturbative description of M-theory via a 
local quantum field theory. 
This involves supersymmetric extension of the bosonic 
Chern-Simons gravity Lagrangian of Chamseddine \cite{Ch1} \cite{Ch2}. 
In section \ref{sec cs}, we show that the bosonic part can be obtained by 
a careful application of the index theorem, following the 
construction of Gilkey \cite{G}. We obtain this term in 
addition to what is already present in the index-theoretic
description of the action, so this suggests that 
the proposed holographic and field-theoretic description of M-theory 
represents a sector within the 
general description via index theory. 

\vspace{3mm}
This paper mainly takes \cite{DFM} as a setting and starting point, applies and physically 
models the mathematical constructions of \cite{DG} \cite{Sh} \cite{BC2} \cite{MP},
 and extends and generalizes the geometric constructions in \cite{Zu}.

\section{Local aspects: The fields on an eleven-manifold with boundary}

We start with a smooth, closed, oriented Riemannian eleven-manifold $Y^{11}$ with a Riemannian metric $g_Y$. Consider $\Omega^p(Y^{11})$, the space of $p$-forms on $Y^{11}$.
The operators relevant for us are the de Rham differential $d: \Omega^p(Y^{11}) \to 
\Omega^{p+1}(Y^{11})$,
the Hodge $\ast$-operator  $\ast: \Omega^p(Y^{11}) \to \Omega^{11-p}(Y^{11})$, 
the co-differential  (the adjoint of $d$) given by
$d^{{}^*}=(-1)^p\ast d \ast: \Omega^p(Y^{11}) \to 
\Omega^{p-1}(Y^{11})$, and the Hodge Laplacian $\Delta=dd^{{}^*} + d^{{}^*} d:
 \Omega^p(Y^{11}) \to \Omega^p(Y^{11})$.
 Define the following subspaces of $\Omega^p(Y^{11})$.
The spaces of exact $p$-forms and co-exact $p$-forms 
on a closed $Y^{11}$ are, respectively,
\begin{eqnarray}
\cE^p (Y^{11})&:=& \{ \omega \in \Omega^p (Y^{11})~:~\omega=d\eta~{\rm for~some~} 
\eta \in \Omega^{p-1}(Y^{11})\}\;,
\label{eq cl}
\\
c\cE^p(Y^{11})&:=& \{ \omega \in \Omega^p (Y^{11})~:~\omega=d^{{}^*} \xi~{\rm for~some~} 
\xi \in \Omega^{p+1}(Y^{11})\}\;.
\label{eq co}
\end{eqnarray}
Furthermore, on a closed $Y^{11}$ there is no distinction between harmonic $p$-fields 
\(
{\rm Harm}^p(Y^{11}):= \{ \omega \in \Omega^p(Y^{11})~:~d\omega=0~{\rm and}~d^{{}^*} \omega=0 \}
\label{eq har}
\)
and harmonic $p$-forms 
\(
\cH^p(Y^{11}):=\{ \omega \in \Omega^p(Y^{11})~:~ \Delta \omega=0\}\;.
\)
The $L^2$ inner product on $\Omega^p(Y^{11})$ is given by
$
\langle \alpha , \beta \rangle_{L^2} = \int_{Y^{11}} \alpha \wedge \ast \beta
$,
for $\alpha, \beta \in \Omega^p(Y^{11})$. 
  When $Y^{11}$ has no boundary, the cohomology of $Y^{11}$ is given, via
the classical de Rham theorem, by the cohomology of the de Rham complex 
 $ 0 \to \Omega^0(Y^{11}) \buildrel{d}\over{\to}
  \Omega^1(Y^{11}) \buildrel{d}\over{\to}
   \Omega^2(Y^{11}) \buildrel{d}\over{\to}
   \cdots
   \buildrel{d}\over{\to}
    \Omega^{10}(Y^{11}) \buildrel{d}\over{\to}
     \Omega^{11}(Y^{11}) \buildrel{d}\over{\to}
0 
 $, that is there is an isomorphism
 $H^p(Y^{11};\R) \cong ({\rm ker}(d): \Omega^p(Y^{11} \to \Omega^{p+1})
 / ({\rm im}(d): \Omega^{p-1}(Y^{11}) \to \Omega^p(Y^{11}))$. 
 The exterior differential commutes with the Laplacian $\Delta$ and hence 
 preserves harmonicity of forms, so that one can build a subcomplex
 $({\rm Harm}^\ast(Y^{11}), d)$ of the de Rham complex, called 
 the {\it harmonic complex} in \cite{CDGM},
 \(
  0 \to {\rm Harm}^0(Y^{11}) \buildrel{d}\over{\to}
  {\rm Harm}^1(Y^{11}) \buildrel{d}\over{\to}
   {\rm Harm}^2(Y^{11}) \buildrel{d}\over{\to}
   \cdots
   \buildrel{d}\over{\to}
    {\rm Harm}^{10}(Y^{11}) \buildrel{d}\over{\to}
     {\rm Harm}^{11}(Y^{11}) \buildrel{d}\over{\to}
0\;. 
 \)
 Being harmonic is equivalent to 
 being closed and co-closed on an eleven-manifold without a boundary, 
 in which case the 
 maps in the harmonic complex are all zero, and by Hodge's theorem 
 $H^p({\rm Harm}^\ast(Y^{11}), d) ={\rm Harm}^p(Y^{11}) \cong H^p(Y^{11} ;\R)$.

 \subsection{Hodge theory on the boundary}
\label{sec hod}
Taking our eleven-manifold to have boundary, several new
phenomena occur. First, the space of harmonic $p$-fields no longer coincide
with the space of harmonic $p$-forms.
Second, the Hodge decomposition theorem is modified as mentioned in the introduction. 
Third, the space of harmonic $p$-fields ${\rm Harm}(Y^{11})$ is infinite dimensional, and 
so is too big to represent cohomology. We will illustrate these ideas in the setting of the 
M-theory fields in what follows.

  \paragraph{Relating harmonic potentials to harmonic field strengths on the boundary.}
 Now consider $Y^{11}$ to have 
 a non-empty boundary $\partial Y^{11}$. 
 In this case the cohomology of the 
 complex $({\rm Harm}^\ast (Y^{11})), d)$ of harmonic forms on $Y^{11}$ is given by the 
 direct sum $H^p({\rm Harm}^\ast(Y^{11}), d)\cong H^p(Y^{11};\R)
 + H^{p-1}(Y^{11};\R)$, for $p=0, 1, \cdots, 11$ \cite{CDGM}. This allows us to characterize 
 the cohomology of $Y^{11}$ in two consecutive degrees in terms of
 the harmonic cohomology in one degree. We are interested in the pairs of degrees
 $(3,4)$ and $(7,8)$, representing the pairs of fields $(C_3, G_4)$ and 
 $(G_7, G_8)$.  For instance, for the first pair we have
 \bea
H^4( {\rm Harm}^\ast(Y^{11}), d) 
=
\frac{\ker(d): {\rm Harm}^4(Y^{11}) \to {\rm Harm}^5(Y^{11})}
{{\rm im}(d):  {\rm Harm}^3(Y^{11}) \to {\rm Harm}^4(Y^{11}) }
&\cong & H^4(Y^{11};\R) + H^3(Y^{11};\R)
 \\
 \left\{ 
G_4 ~|~ dG_4=0, \Delta G_4=0, \nexists ~C_3 {\rm ~with~} \Delta C_3=0
{\rm ~so ~that~} G_4 \neq dC_3
\right\}
&=&
\left\{
G_4 ~|~dG_4=0, G_4\neq dC_3
\right\}
\\
&+&
\left\{
C_3 ~|~dC_3=0, C_3\neq dB_2
\right\}
\;,
\eea
where $B_2$ is 2-form on $Y^{11}$.  
 Then we have 
  
\begin{proposition}
Consider M-theory on a manifold with a boundary.
Harmonic closed non-exact field strengths $G_4$ are equivalent to 
closed non-exact $G_4$ together with closed non-exact C-field 
$C_3$. 
\end{proposition}
Note that in order to deal with both the equations of motion and the Bianchi 
 identities, we can similarly look at the subcomplex of harmonic forms 
 on $Y^{11}$ with differential $d^{{}^*}$, with the obvious changes.

\vspace{3mm}
Let $i: M^{10}=\partial Y^{11} \to Y^{11}$ be the inclusion map. 
Let $d_\partial$, $\ast_\partial$, $d^{{}^*}_\partial$ and $\Delta_\partial$, respectively,
 denote the exterior derivative,
Hodge star, co-differential and Laplacian on the closed Riemannian ten-manifold $M^{10}$. 
The relative cohomology group is defined for smooth $p$-forms 
$\omega$ whose pullback $i^*\omega$ to $M^{10}$ is zero. Define relative $p$-forms by
$
\Omega^p(Y^{11}, \partial Y^{11}):=\{ \omega \in \Omega^p(Y^{11}) : i^* \omega=0 \}
$, in which the subspaces of closed and exact relative $p$-forms are given by
the conditions $d\omega=0$ and $\omega=d\eta$ for some $\eta\in \Omega^{p-1}(Y^{11}, \partial Y^{11})$,
respectively, and whose quotient is the relative de Rham cohomology.

\paragraph{Boundary conditions on harmonic forms.}
We would like to pull back harmonic forms on $Y^{11}$ to harmonic forms on 
the boundary $\partial Y^{11}$. The simplest question to ask 
is whether such a pullback is zero. We start on $Y^{11}$
with a C-field $C_3$, which is closed and co-closed, that is $dC_3=0$ and
$d^{{}^*} C_3=0$. Then consider the pullback $C'_3=i^* C_3$. If we take this form to 
be zero then $C_3$ itself must be zero. This generalizes to other forms as well,
so we get the same conclusion for $G_4, G_7$ and $G_8$, using the general
results in \cite{CDGM}. Therefore, 

\begin{proposition}
Let ${\cal C}$ stand for any of the fields $C_3, G_4, G_7$ or $G_8$ on $Y^{11}$
which is closed and co-closed: $d{\cal C}=0$, $d^{{}^*} {\cal C}=0$. 
Let ${\cal C}'=i^* {\cal C}$ be the pullback of ${\cal C}$ to the boundary 
$\partial Y^{11}$. If this pullback is zero then the original  field is identically zero:
${\cal C}'=0$ implies ${\cal C}=0$. In particular, for a nontrivial $G_4$ which satisfies the equations
of motion and the Bianchi identity in the limit $\ell_p \to 0$, we cannot take 
the pullback to be zero. 
\end{proposition}
Therefore, we work with nonzero pullbacks to the boundary.


\paragraph{Dirichlet vs. Neumann forms.}
The boundary condition $i^*\omega=0$ appearing in the definition of relative $p$-forms 
is called a Dirichlet boundary condition. A form $\omega$ satisfying the Dirichlet boundary 
condition can be thought of as being normal to the boundary; for any $x\in M^{10}$ and 
$v_1, \cdots, v_p\in T_xY^{11}$, the form $\omega( v_1, \cdots, v_p)$ is not zero only if one of the 
vectors has a nontrivial component in the direction of the inward-pointing unit normal vector 
$\mathfrak{i}_n$.
For $x\in M^{10}$, let pr$: T_xY^{11}\to T_xM^{10}$ be the orthogonal projection. 
A $p$-form $\omega$ on $Y^{11}$ is tangent to the boundary if 
$\omega(v_1, \cdots, v_p)= \omega({\rm pr}(v_1), \cdots, {\rm pr}(v_p))$ or, equivalently, 
if the contraction $\mathfrak{i}_n \omega$ is zero, that is 
\(
0=\ast_\partial \mathfrak{i}_n \omega= i^* \ast \omega\;.
\)
This is the Neumann boundary condition. In \cite{LM} the authors consider
$\frak{i}_n G'_4$ the 3-form that comes from contracting the 
M-theory G-flux in the bulk $Y^{11}$ with the normal unit 
vector field to the boundary. 
Now the Hodge $\ast$ exchanges Dirichlet forms with Neumann forms, so that
if $\omega \in \Omega^p(Y^{11})$ satisfies the Neumann condition then 
the dual $\ast \omega \in \Omega^{11-p}(Y^{11})$ satisfies the Dirichlet condition, and 
vice versa. We immediately have 

\begin{proposition} 
If $G_4$ (or $C_3$) is a Neumann form then $\ast G_4$ 
(or $\ast C_3$) is a Dirichlet form and vice versa.  Hence the Dirichlet condition cannot be 
applied both to the C-field and its dual. Consequently, both fields cannot take values in  
relative cohomology.
\end{proposition}

Let the subscripts $N$ and $D$ denote Neumann and Dirichlet boundary conditions 
respectively. Consider the counterparts of the spaces 
\eqref{eq cl}, \eqref{eq co}, and \eqref{eq har} in the presence of a boundary
\begin{eqnarray}
\cE^p_D(Y^{11})&:=& \{ \omega \in \Omega^p (Y^{11})~:~\omega=d\eta~{\rm for~some~} 
\eta \in \Omega^{p-1}(Y^{11}) {\rm ~where~} i^*\eta=0\}\;,
\\
c\cE^p_N(Y^{11})&:=& \{ \omega \in \Omega^p (Y^{11})~:~\omega=d^{{}^*} \xi~{\rm for~some~} 
\xi \in \Omega^{p+1}(Y^{11}) {\rm ~where~} i^* \ast \xi=0\}\;,
\\
{\rm Harm}^p_D(Y^{11})&:=& \{ \omega \in \Omega^p~:~d\omega=0, d^{{}^*} \omega=0, {\rm~and~}
i^*\omega=0
 \}\;,
 \\
 {\rm Harm}^p_N(Y^{11})&:=& \{ \omega \in \Omega^p~:~d\omega=0, d^{{}^*} \omega=0, {\rm~and~}
i^* \ast \omega=0 \}\;.
\end{eqnarray}
Note that the boundary conditions apply to the primitive of $\omega$ in the first two spaces
above, while they apply to $\omega$ itself in the last two. 
If $\omega \in \cE_D^p(Y^{11})$ then $\omega= d\eta$ for some Dirichlet form
$\eta \in \Omega^{p-1}(Y^{11})$ and $i^* d\eta= d_\partial i^* \eta=0$, so 
$\cE_D^p(Y^{11})$  is the space of {\it relatively exact $p$-forms}. 
If $\omega \in c\cE_N^p(Y^{11})$, then 
$\omega= d^{{}^*} \xi$ for some Neumann form $\xi \in \Omega^{p+1}(Y^{11})$ and 
$i^* \ast \xi= (-1)^{p+1} i^* d\ast \xi = (-1)^{p+1} d_\partial i^* \ast \xi =0$, so forms 
in $c\cE_N^p(Y^{11})$ satisfy the Neumann boundary condition. 
The Hodge $\ast$ operator then takes ${\rm Harm}_D^{11-p}(Y^{11})$ to
${\rm Harm}_N^p(Y^{11})$.

\paragraph{$L^2$-decompositions and the Hodge-Morrey-Friedrics (HMF) decomposition.}
We would like to consider the kinetic terms for the fields (cf. equation \eqref{exp ac}), 
and hence we are interested
in $L^2$ (square integrable) expressions. 
Let $\oplus_{L^2}$ denote orthogonal sum while 
$\oplus$ denotes direct sum. 
The Hodge-Morrey-Friedrichs decomposition theorem in our case is
the $L^2$-orthogonal direct sum (see \cite{Sh})
\begin{eqnarray}
\Omega^p(Y^{11})&=& c\cE_N^p(Y^{11}) \oplus_{L^2} {\rm Harm}_N^p(Y^{11}) 
\oplus_{L^2} \cE^p(Y^{11}) \cap {\rm Harm}^p(Y^{11}) \oplus_{L^2}
\cE_D^p(Y^{11})
\nonumber\\
&=& c\cE_N^p(Y^{11}) \oplus_{L^2} c\cE^p(Y^{11}) \cap {\rm Harm}^p(Y^{11})
\oplus_{L^2} {\rm Harm}_N^p(Y^{11}) \oplus_{L^2} \cE_D^p(Y^{11})\;,
\label{hmf eq}
\end{eqnarray}
with the absolute and relative cohomology groups given by
\(
H^p(Y^{11};\R) \cong {\rm Harm}_N^p(Y^{11})\;, \qquad \qquad
H^p(Y^{11}, \partial Y^{11};\R) \cong 
{\rm Harm}_D^p(Y^{11})\;.
\)
Applying to the field strength $G_4$, and denoting the space of these fields
by $\{G_4\}$, we get 
\begin{proposition}
$(i)$ The space of C-fields in M-theory with a boundary, in the limit 
$\ell_p \to 0$, decomposes into four orthogonal 
spaces 
$$
\{ G_4\}= 
\left\{
\begin{array}{c}
G_4=\ast dC_6
\\
i^*C_6=0
\end{array}
\right\}
\oplus_{L^2}
\left\{
\begin{array}{c}
dG_4=0
\\
d^{{}^*} G_4=0
\\
i^*\ast G_4=0
\end{array}
\right\}
\oplus_{L^2}
\left\{
\begin{array}{c}
G_4=dC_3
\\
d^{{}^*} G_4=0
\end{array}
\right\}
\oplus_{L^2}
\left\{
\begin{array}{c}
G_4=dC_3
\\
i^*C_3=0
\end{array}
\right\}
$$
$(ii)$ Denote the above $L^2$ summands in  
$(i)$ 
as type 1, 2, 3, and 4, with corresponding fields 
$G_4^{(1)}$, $G_4^{(2)}$, $G_4^{(3)}$, and $G_4^{(4)}$, 
respectively. Then the kinetic term for the C-field decomposes as
$$
\langle G_4, G_4 \rangle_{L^2}
=
\langle G_4^{(1)}, G_4^{(1)} \rangle_{L^2}
+
\langle G_4^{(2)}, G_4^{(2)} \rangle_{L^2}
+
\langle G_4^{(3)}, G_4^{(3)} \rangle_{L^2}
+
\langle G_4^{(4)}, G_4^{(4)} \rangle_{L^2}\;.
$$
\label{Thm dec}
\end{proposition}

\vspace{-3mm}
Consider the kinetic term for the C-field as in Proposition \ref{Thm dec}.
When $Y^{11}$ has no boundary the variational principle gives 
$d \ast G_4=0$ for the equation of motion as in \eqref{EOM}.
Now in the presence of a boundary $M^{10}=\partial Y^{11}$, Green's formula 
gives, for $\alpha \in \Omega^{p-1} (Y^{11})$ and $\beta \in \Omega^p(Y^{11})$, 
$
\langle d\alpha, \beta \rangle_{L^2} -
\langle \alpha, d^{{}^*} \beta \rangle_{L^2} 
= \int_{M^{10}} i^* \alpha \wedge i^* \ast \beta$.
Therefore, for the C-field we get (we use prime $\Xi'$ for boundary fields)
\begin{lemma}
When $\partial Y^{11} \neq \emptyset$, we have 
$
\int_{Y^{11}} dC_3 \wedge \ast G_4 - \int_{Y^{11}} C_3 \wedge d \ast G_4
= \int_{\partial Y^{11}} C'_3 \wedge G'_7$,
where $C'_3:=i^*C_3$ and $G'_7:=i^* \ast G_4$.
\label{lem c3g7}
\end{lemma}
Now a more precise statement than 
the one given in the introduction is the consequence of 
the Hodge-Morrey-Friedrichs decomposition theorem 
$
{\rm Harm}_N^p (Y^{11}) \cap {\rm Harm}_D^p(Y^{11})=\{0\}$.
For the C-field this means that in the limit $\ell_p \to 0$
\(
\left\{
dG_4=0\;,
d^{{}^*} G_4=0\;,
i^* \ast G_4=0
\right\}
\cap
\left\{
dG_4=0\;,
d^{{}^*} G_4=0\;,
i^*G_4=0
\right\}
=\{ 0\}\;.
\)
So now we ask whether both of the above spaces can appear in the 
orthogonal decomposition of a general field strength $G_4 \in \Omega^4(Y^{11})$.
A consequence of the the HMF theorem is 
the DeTurck-Gluck decomposition \cite{DG}, in which the two outer terms in 
\eqref{hmf eq} are the same, but the two 
inner terms are replaced by 
$\cE^p(Y^{11}) \cap c\cE^p(Y^{11}) \oplus
({\rm Harm}_N^p(Y^{11}) + {\rm Harm}_D^p(Y^{11}))$.
For $G_4$ we then have the following alternative to the  
decomposition in Proposition \ref{Thm dec}
\(
\left\{
\hspace{-2mm}
\begin{array}{c}
G_4=\ast dC_6
\\
i^*C_6=0
\end{array}
\hspace{-2mm}
\right\}
\oplus_{L^2}
\left\{
\hspace{-2mm}
\begin{array}{c}
G_4= dC_3
\\
G_4=\ast dC_6
\end{array}
\hspace{-2mm}
\right\}
\oplus_{L^2}
\left[
\left\{
\begin{array}{c}
dG_4=0
\\
d^{{}^*} G_4=0
\\
i^* \ast G_4=0
\end{array}
\right\}
\oplus
\left\{
\begin{array}{c}
dG_4=0
\\
d^{{}^*} G_4=0
\\
i^* G_4=0
\end{array}
\right\}
\right]
\oplus_{L^2}
\left\{
\hspace{-2mm}
\begin{array}{c}
G_4=dC_3
\\
i^*C_3=0
\end{array}
\hspace{-2mm}
\right\}
\label{eq nonor}
\)
Here we will have a decomposition of the kinetic term of the C-field as
in part (ii) of Proposition \ref{Thm dec}, taking into account the non-orthogonal direct sum in 
\eqref{eq nonor}.

\vspace{3mm}
The reason for the non-orthongonality, as explained more generally in \cite{Sh}, 
is the fact that some of the 
cohomology of $Y^{11}$ comes from the interior of $Y^{11}$ and some comes
from the boundary $\partial Y^{11}$. 
First, in absolute cohomology the interior part is 
${\rm ker}(i^*: H^p(Y^{11};\R) \to H^p(\partial Y^{11};\R))$. 
This is the condition imposed in \cite{DFM} (see the introduction).
Since $H^p(Y^{11};\R) \cong {\rm Harm}^p(Y^{11})$
then the boundary portion is the subspace of the harmonic 
Neumann fields which pull back to zero in the cohomology of the boundary, 
\(
\cE_\partial^p(\partial Y^{11}) \cap {\rm Harm}_N^p(Y^{11}):=
\left\{ 
\omega \in {\rm Harm}_N^p(Y^{11})~:~i^* \omega=d\varphi {\rm ~for~some~} 
\varphi \in \Omega^{p-1}(\partial Y^{11})
\right\}\;.
\label{eq A}
\)
For the field strength $G_4$ of the C-field we have
$$
\cE_\partial^4(\partial Y^{11}) \cap {\rm Harm}^4_N(Y^{11}):=
\left\{ 
G_4 
~|~
dG_4=0\;,
d^{{}^*} G_4=0\;,
i^* \ast G_4=0
~:~
i^* G_4= dC'_3
{\rm ~for~some~} 
C'_3 \in \Omega^{3}(\partial Y^{11})
\right\}\;.
$$
Second, 
for relative cohomology, let $j: Y^{11}=(Y^{11}, \emptyset) \to (Y^{11}, \partial Y^{11})$ be 
the inclusion. 
The Hodge $\ast$ operator exchanges the 
space ${\rm Harm}_D^4(Y^{11})$ with ${\rm Harm}_N^7(Y^{11})$ and 
 ${\rm Harm}_D^7(Y^{11})$ with ${\rm Harm}_N^4(Y^{11})$.
The Hodge $\ast$ also exchanges the 
boundary subspace 
$\cE^{7}(Y^{11})\cap {\rm Harm}_D^{7}(Y^{11})$ with the boundary subspace
$c \cE^4(Y^{11}) \cap {\rm Harm}_N^4(Y^{11})$. For the C-field, we have
the effect as the exchange
\(
\left\{
\begin{array}{cc}
G_4=dC_3, & dG_4=0
\\
d^{{}^*} G_4=0, & i^* \ast G_4=0
\end{array}
\right\}
~\buildrel{\ast}\over{\longleftrightarrow}~
\left\{
\begin{array}{cc}
G_7=\ast dC_3, & dG_7=0
\\
d^{{}^*} G_7=0, & i^* G_7=0
\end{array}
\right\}\;.
\label{eq lr}
\)
Similar statements can be deduced for $G_4$ replaced with $G_7$ and $C_3$ replaced
with $C_6$.
The portion in relative cohomology coming from the boundary is 
formed out of 
those Dirichlet fields which are exact, $\cE^p(Y^{11})\cap {\rm Harm}_D^p(Y^{11})$, while the portion
coming from the interior is the subspace
$
c\cE_\partial^p(\partial Y^{11}) \cap {\rm Harm}_D^p=\left\{
\omega \in {\rm Harm}_D^p(Y^{11})~:~i^* \ast \omega = d\psi \right\}$
for~some $\psi \in \Omega^{10-p}(\partial Y^{11})$.
This the same as the space on the left in \eqref{eq lr} except that 
$C_3 \in \Omega^3(Y^{11})$ is replaced with $C'_3\in \Omega^3(\partial Y^{11})$.
So the spaces ${\rm Harm}_N^p(Y^{11})$ and ${\rm Harm}_D^p(Y^{11})$ admit the
$L^2$-orthogonal decomposition into interior and boundary subspaces \cite{DG}
\bea
{\rm Harm}_D^4(Y^{11}) &=& \cE^4(Y^{11}) \cap {\rm Harm}_D^4(Y^{11}) 
\oplus c\cE_\partial^4(\partial Y^{11})  \cap {\rm Harm}_D^4(Y^{11})\;,
\\
{\rm Harm}_N^4(Y^{11}) &=& c\cE^4(Y^{11}) \cap {\rm Harm}_N^4(Y^{11}) 
\oplus \cE_\partial^4(\partial Y^{11}) \cap  {\rm Harm}_N^4(Y^{11})\;.
\eea
For the C-field the first of the two expressions gives
the $L^2$-orthogonal decomposition 
\(
\left\{
\hspace{-2mm}
\begin{array}{cc}
G_4 & dG_4=0
\\
d^{{}^*} G_4=0, & i^* G_4=0
\end{array}
\hspace{-2mm}
\right\}
=
\left\{
\hspace{-2mm}
\begin{array}{cc}
G_4=dC_3, & dG_4=0
\\
d^{{}^*} G_4=0, & i^* G_4=0
\end{array}
\hspace{-2mm}
\right\}
\oplus_{L^2}
\left\{
\hspace{-2mm}
\begin{array}{cl}
G_4, & dG_4=0
\\
d^{{}^*} G_4=0, & \hspace{-2mm} i^* G_4=dC'_3, ~C'_3\in \Omega^3(\partial Y^{11})
\end{array}
\hspace{-2mm}
\right\}
\;,
\label{eq ooo}
\)
while the second gives 
\(
\left\{
\hspace{-2mm}
\begin{array}{cc}
G_4 & \hspace{-3mm}dG_4=0
\\
d^{{}^*} G_4=0, & \hspace{-2mm}i^* \ast G_4=0
\end{array}
\hspace{-2mm}
\right\}
=
\left\{
\hspace{-2mm}
\begin{array}{cc}
G_4=\ast dC_6, & \hspace{-2mm}dG_4=0
\\
d^{{}^*} G_4=0, & \hspace{-2mm}i^* \ast G_4=0
\end{array}
 \hspace{-2mm}
\right\}
\oplus_{L^2}
\left\{ \hspace{-2mm}
\begin{array}{cl}
G_4, & dG_4=0
\\
d^{{}^*} G_4=0, &\hspace{-2mm} i^* \ast G_4=dC'_6, ~C'_6\in \Omega^6(\partial Y^{11})
\end{array}
\hspace{-2mm}\right\}
\;.
\label{eq ppp}
\)
Let us illustrate equation \eqref{eq ppp}; write the two summands on the right
hand side as $Q \oplus R$. Letting $d^{{}^*} \xi_5$ (or $\ast dC_6$) $\in Q$ and $\tilde{G}_4 \in R$
we have $i^* \tilde{G}_4=dC'_3$ for some $C'_3 \in \Omega^3(\partial Y^{11})$. Then 
\begin{eqnarray}
\langle \tilde{G}_4, d^{{}^*} \xi_5 \rangle_{L^2} &=&
\langle d \tilde{G}_4, \xi_5 \rangle_{L^2}
- 
\int_{\partial Y^{11}} i^* \tilde{G}_4 \wedge i^* \ast \xi_5 
\nonumber\\
&=& - \int_{\partial Y^{11}} dC'_3 \wedge i^* \ast \xi_5
\nonumber\\
&=& 
\langle d\tilde{C}_3, d^{{}^*} \xi_5 \rangle_{L^2} \hspace{5cm}
\tilde{C}_3 \in \Omega^3(Y^{11})~{\rm extension~of~} C'_3 {\rm ~to~} Y^{11}
\nonumber\\
&=&
\langle \tilde{C}_3, d^{{}^*} d^{{}^*} \xi_5 \rangle_{L^2} + \int_{Y^{11}}
i^* \tilde{C}_3 \wedge i^* \ast d^{{}^*} \xi_5 \hspace{1.65cm} {\rm Green's~Theorem}
\nonumber\\
&=& 0 \hspace{6.7cm} i^* \ast d^{{}^*} \xi_5=0 {\rm ~since~} d^{{}^*} \xi_5 {\rm ~is~Neumann.} 
\nonumber
\end{eqnarray}
Write the decompositions \eqref{eq ooo} and \eqref{eq ppp} schematically as
\(
\{ G_4^D\}= \{ G_4^{D,e}\} \oplus_{L^2} \{ G_4^{D,c}\}\;,
\qquad \qquad
\{ G_4^N\}= \{ G_4^{N,c}\} \oplus_{L^2} \{ G_4^{N,e}\}\;,
\)
where the superscripts $D$ and $N$ refer to Dirichlet and Neumann, while
the extra superscripts $e$ and $c$ refer to exact and co-exact, respectively. 
We summarize the above discussion with 
\begin{proposition}
The field strength splits into Dirichlet and Neumann forms $G_4^D$ and 
$G_4^N$, whose 
$L^2$-inner product  
decomposes according to 
Dirichlet and Neumann boundary conditions and to exactness and co-exactness
on the boundary as 
\bea
\langle G_4^D, G_4^D \rangle_{L^2} &=& \langle G_4^{D,e}, G_4^{D,e} \rangle_{L^2} 
\oplus_{L^2} \langle G_4^{D,c}, G_4^{D,c} \rangle_{L^2}\;,
\\
\langle G_4^N, G_4^N \rangle_{L^2} &=& \langle G_4^{N,e}, G_4^{N,e} \rangle_{L^2} 
\oplus_{L^2} \langle G_4^{N,c}, G_4^{N,c} \rangle_{L^2}\;.
\eea 
\label{prop pure}
\end{proposition}
\vspace{-3mm}
This is a refinement of the split of $G_4$ into $G_4^D$ and $G_4^N$, 
for instance in \cite{LM}.
We will consider ``mixing" in section \ref{sec PD}. 

\paragraph{Integral forms.} The partition function of the C-field
requires integrating over the space of harmonic forms as well as
summing over torsion fields. We have considered the former in 
a lot of detail so far, so we now provide remarks about including 
the latter. 
Denote by ${\rm Harm}_{N, \Z}^p(M)$ the image in 
${\rm Harm}_{N}^p(M)$ of the integer lattice 
$H^p_\Z(M;\R)$ of $H^p(M;\R)$ under the isomorphism
$H^p(M;\R) \cong {\rm Harm}_N^p(M)$.
A form $\alpha$ is in ${\rm Harm}_{N, \Z}^p(M)$
if and only if $\int_S \alpha \in \Z$ for any singular
$p$-cycle $S$ of $M$. 
Similarly, 
Denote by ${\rm Harm}_{D, \Z}^p(M)$ the image in 
${\rm Harm}_{D}^p(M)$ of the integer lattice 
$H^p_\Z(M, \partial M;\R)$ of $H^p(M,\partial M;\R)$ under the isomorphism
$H^p(M, \partial M;\R) \cong {\rm Harm}_D^p(M)$.
A form $\alpha$ is in ${\rm Harm}_{D, \Z}^p(M)$
if and only if $\int_S \alpha \in \Z$ for any relative singular
$p$-cycle $S$ of $M$. With these notions, the above discussions can be extended
(but we do not need that explicitly here).

\subsection{Duality on the boundary and boundary value problems}
\label{sec bvp}
We seek to characterize the resulting duality between the fields pulled back
to the boundary $\partial Y^{11}$, starting with Hodge duality on $Y^{11}$.
The expression in Lemma \ref{lem c3g7} suggests that we look at $C'_3$ and 
$G'_7$. 

\paragraph{Dirichlet-to-Neumann map on the forms.}
There is an operator that takes care of Hodge duality on the boundary and 
which treats the field and its dual at the same time
\cite{JL} \cite{SS}.
Define the Dirichlet-to-Neumann (D-to-N) map $\Lambda_p: \Omega^p(\partial Y^{11})
\to \Omega^{10-p}(\partial Y^{11})$ as follows 
\cite{BS} \cite{JL} \cite{Sc}. If $\varphi \in \Omega^p(\partial Y^{11})$
is a smooth $p$-form on the boundary
the define $\Lambda_p \varphi:= i^* \ast d\omega$, which is independent of the 
choice of $\omega$ due to the presence of the exterior derivative. 
Hence for the boundary fields $C'_3$ and $G'_7$ we have 
$\Lambda_3 C'_3:= i^* \ast dC_3$ and $\Lambda_7 G'_7:= i^* \ast dG_7$ with
$C_3$ and $G_7$ a three- and seven-form, respectively, on $Y^{11}$.
Then the boundary value problem 
\(
\Delta \omega=0, \quad i^* \omega=\varphi, {\rm ~~and~~} i^*d^{{}^*} \omega=0
\label{bvp1}
\)
has a unique solution up to the addition of an arbitrary harmonic 
Dirichlet field $\lambda \in {\rm Harm}_D^p(Y^{11})$ \cite{Sc}. 
From \cite{BS}, $d\omega \in {\rm Harm}^{p+1}(Y^{11})$ and $d^{{}^*} \omega=0$, so that
\eqref{bvp1} is equivalent to 
the boundary value problem 
\(
\Delta \omega=0, \quad i^* \omega=\varphi, {\rm ~~and~~} d^{{}^*} \omega=0\;.
\label{bvp12}
\)
We then have, in our case, the following two BVPs for the two boundary fields $C'_3$ and 
$G'_7$
\begin{eqnarray}
{\rm BVP}1~&:&~ \Delta C_3=0\;, \qquad i^*C_3=C'_3\;, \qquad d^{{}^*} C_3=0\;,
\\
{\rm BVP}2~&:&~\Delta G_7=0\;, \qquad i^*G_7=G'_7\;, ~~\quad d^{{}^*} G_7=0\;.
\end{eqnarray}
The kernel and image of the D-to-N operator are given by 
$
i^* {\rm Harm}^p(Y^{11})= {\rm ker} \Lambda_p = {\rm im} \Lambda_{10-p}$,
which gives, for a 3-form, $i^* {\rm Harm}^3(Y^{11})={\rm ker} \Lambda_3= {\rm im} \Lambda_7$.
In fact, from \cite{Sh}, this kernel has a direct sum decomposition 
$
{\rm ker} \Lambda_p =i^* {\rm Harm}^p(Y^{11})=
i^* c\cE^p(Y^{11}) \cap {\rm Harm}^p(Y^{11}) + 
\cE^p(\partial Y^{11})$,
so that 
${\rm ker} \Lambda_p/\cE^p(\partial Y^{11})\cong  c\cE^p(Y^{11}) \cap {\rm Harm}^p(Y^{11})$,
with dimension equal to that of the boundary subspace of $H^p(Y^{11};\R)$. 
For our two fields on the boundary we have
\begin{eqnarray}
{\rm ker} \Lambda_3&=& \left\{
C'_3=i^* C_3 ~|~ C_3=d^{{}^*} G_4\;, ~dC_3=0\;,  ~d^{{}^*} C_3=0
\right\}
\oplus
\{ 
C'_3=dB'_2
\}\;,
\\
{\rm ker} \Lambda_7&=& \left\{
G'_7=i^* G_7 ~|~ G_7= d^{{}^*} G_8\;, ~dG_7=0\;,  ~d^{{}^*} G_7=0
\right\}
\oplus
\{ 
G'_7=dB'_6
\}\;.
\end{eqnarray}
This D-to-N operator 
can treat the field and its dual in a unified way by considering
\(
\Pi : \Omega^k (\partial Y^{11}) \times \Omega^{11-k}(\partial Y^{11})
\longrightarrow 
\Omega^{10-k} (\partial Y^{11}) \times \Omega^{k-1}(\partial Y^{11})
\)
defined by 
$
\Pi 
\binom{\varphi}{\psi}
=
\binom{i^*\ast d\omega}{i^* d^{{}^*} \omega}
$,
where $\omega \in \Omega^k(Y^{11})$ is the solution of the boundary value problem 
$
\{\Delta \omega=0,  i^* \omega= \varphi, i^* \ast \omega = \psi\}$. 
This is best described by splitting into two linear operators $(\Phi, \Psi)$
\cite{SS}
as $\Pi=\binom{\Phi~~,~~ (-1)^k \Psi}{\Psi~, ~(-1)^{k+1}\Phi}$
on $\Omega^k(\partial Y^{11}) \times \Omega^{11-k}(\partial Y^{11})$,
with
$
\Phi: \Omega^k(\partial Y^{11}) \to \Omega^{10-k}(\partial Y^{11})$
and
$\Psi: \Omega^k(\partial Y^{11}) \to \Omega^{k-1}(\partial Y^{11})$
defined by the expressions 
$
\Phi \varphi= i^* \ast d\omega$ and  
$\Psi \varphi = i^* d^{{}^*} \omega$,
where $\omega\in \Omega^k(Y^{11})$ is now a solution to the boundary value problem
$\{
\Delta \omega=0, i^* \omega= \varphi, i^* \ast \omega = 0\}
$.
We are interested in the pair $k=(3, 7)$, which is equivalent to the pair $k=(4, 8)$ 
--
the effect will be simply an exchange of factors; for instance, for $k=3$ we will 
have $\Pi: \Omega^3(\partial Y^{11}) \times \Omega^8(\partial Y^{11})
\to \Omega^7(\partial Y^{11}) \times \Omega^2(\partial Y^{11})$, while 
for $k=8$ we have 
$\Pi: \Omega^8(\partial Y^{11}) \times \Omega^3(\partial Y^{11})
\to \Omega^2(\partial Y^{11}) \times \Omega^7(\partial Y^{11})$.

\vspace{3mm}
Let $b_k(Y^{11})={\rm dim}H^k(Y^{11};\R)$ be the $k$th 
Betti number of $Y^{11}$. Then, from \cite{SS}, $b_k(Y^{11})= {\rm dim}({\rm ker}(\Phi))$
and the kernel of the operator 
$\Phi_k: \Omega^k(\partial Y^{11}) \to \Omega^{10-k}(\partial Y^{11})$
consists of the boundary traces of harmonic Neumann fields, i.e. 
${\rm ker}(\Phi_k)= i^* {\rm Harm}_N^k(Y^{11})$.
For $C'_3$ and $G'_7$ we have

\begin{theorem} 
$(i)$ The (solvable) boundary value problem which involves duality on the boundary is given by
\begin{eqnarray}
 \Pi \left( 
\begin{array}{cc}
C'_3  \\
G'_8  
\end{array}
\right)
=
 \left( 
\begin{array}{cc}
i^* \ast dC_3  \\
i^* d^{{}^*} C_3  
\end{array}
\right)\;,
\Phi_3 C'_3=i^* \ast dC_3, \Psi_3 C'_3=i^* d^{{}^*} C_3,
C_3 
{\rm ~is~a~solution~ to~BVP~} 
\nonumber\\
\{ 
\Delta C_3=0, i^* C_3=C'_3, i^* \ast C_3 =G_8
\}  
\nonumber
\end{eqnarray}
and similarly for $G_7$ with $\Phi_7 G'_7=i^* \ast d G_7$
and $\Psi_7 G'_7= i^* d^{{}^*} G_7$.

\noindent $(ii)$ The Betti numbers of $Y^{11}$ are given in terms of the D-to-N operators as 
\bea
b_3(Y^{11})&=&{\rm dim}~{\rm ker}(\Phi_3)
=
{\rm dim}(i^* {\rm Harm}_N^3(Y^{11}))=
{\rm dim} \{
C'_3~|~C_3'=i^* C_3, dC_3=0, d^{{}^*} C_3=0, i^* \ast C_3=0
\}\;,
\\
b_7(Y^{11})&=&{\rm dim}~{\rm ker}(\Phi_7)
=
{\rm dim}(i^* {\rm Harm}_N^7(Y^{11}))=
{\rm dim} \{
G'_7~|~G'_7=i^*G_7, dG_7=0, d^{{}^*} G_7=0, i^* \ast G_7=0
\}\;.
\eea

\end{theorem}


\subsection{Mixing Dirichlet with Neumann and Poincar\'e duality angles}
\label{sec PD}

We have seen how the absolute harmonic forms and the relative harmonic
forms have no elements in common except for 0. We have also seen that
the interior parts of the absolute and the relative cohomology have 
elements in common. We now characterize these elements making use of results from 
\cite{Sh}.
To that end, we consider the Poincar\'e duality angles, which 
are the principal angles between the following two interior 
subspaces (cf. equation \eqref{eq A})
\bea
{\cal A}:=\cE_\partial^3 (\partial Y^{11}) \cap {\rm Harm}_N^3(Y^{11})=
\left\{ 
C_3 
~|~
dC_3=0, 
d^{{}^*} C_3=0,
i^* \ast C_3=0
~:~
i^* C_3= dB'_2
{\rm ~for~}
B'_2 \in \Omega^{2}(\partial Y^{11})
\right\}\;,
\nonumber\\
{\cal B}:=c\cE_\partial^3 (\partial Y^{11}) \cap {\rm Harm}_D^3(Y^{11})=
\left\{ 
C_3 
~|~
dC_3=0, 
d^{{}^*} C_3=0,
i^*  C_3=0
~:~
i^* \ast C_3= dG'_7
{\rm ~for~}
G'_7 \in \Omega^{7}(\partial Y^{11})
\right\}\;.
\eea
Following the general construction in \cite{Sh}, 
let ${\sf proj}_D: {\rm Harm}_N^3(Y^{11})  \to {\rm Harm}_D^3(Y^{11})$ 
be the orthogonal projection onto the space of Dirichlet fields.
This takes a closed and co-closed $C_3$ whose Hodge dual pulls back to 
zero on the boundary to one which is closed and co-closed and itself
 pulls back to zero on the boundary. 
Then 
${\sf proj}_D {\cal A}={\cal B}$. 
The nonzero singular values of ${\sf proj}_D$, that is the square roots of the eigenvalues of the 
nonnegative adjoint operator ${\sf proj}_D^* \circ {\sf proj}_D$, are given by $\cos \theta$
and so define the Poincar\'e duality angle $\theta$. 
This is also given by the nonzero singular values of ${\sf proj}_N:  {\rm Harm}_D^3(Y^{11})\to {\rm Harm}_N^3(Y^{11})$ onto the space of Neumann fields.
Then $\cos^2 \theta$ are the nonzero eigenvalues of the compositions
${\sf proj}_N \circ {\sf proj}_D: {\rm Harm}_N^3(Y^{11})\to {\rm Harm}_N^3(Y^{11})$ 
and ${\sf proj}_D \circ {\sf proj}_N: {\rm Harm}_D^3(Y^{11})\to {\rm Harm}_D^3(Y^{11})$.  

\vspace{3mm}
Let $T$ denote the Hilbert transform \cite{BS} 
defined as $T:=d_\partial \Lambda^{-1}$ 
and which is well-defined on the subset of forms on $\partial Y^{11}$ given by 
$i^* {\rm Harm}^p(Y^{11})={\rm im} \Lambda_{7}$.
Let $C^N_3 \in {\rm Harm}_N^3(Y^{11})$ and 
${\sf proj}_D C^N_3 =  C^D_3\in {\rm Harm}_D^3(Y^{11})$ be the orthogonal projection 
of $C_3^N$ onto ${\rm Harm}_D^3(Y^{11})$.
On the other hand, let $C_3^D \in {\rm Harm}_D^3(Y^{11})$ and 
${\sf proj}_N C_3^D = C_3^N \in {\rm Harm}_N^3(Y^{11})$ be the orthogonal projection 
of $C_3^D$ onto ${\rm Harm}_N^3(Y^{11})$.
Then, using \cite{Sh},
\(
T i^* C_3^N =  i^* \ast C_3^D\;, \qquad
T i^* \ast C_3^D = - i^* C_3^N\;.
\)
There is connection between Dirichlet-to-Neumann map and the Poincar\'e duality angles.
Consider the restriction $\widetilde{T}$ of the Hilbert transform $T$ 
to the pullback of Neumann harmonic three-forms
\(
i^* {\rm Harm}_N^3(Y^{11})=
\{
C'_3 ~|~ C'_3=i^* C_3, dC_3=0, d^{{}^*}C_3=0, i^* \ast C_3=0
\}\;.
\)
 Then \cite{Sh} the quantities 
$-\cos^2 \theta$ are the nonzero eigenvalues of $\widetilde{T}^2$. 
We can similarly consider $G_4$ in place of $C_3$ for which
similar results as above hold. 
\begin{proposition} $(i)$ Let $G_4^N \in {\rm Harm}_N^4(Y^{11})$
and ${\sf proj}_D G_4^N = G_4^D \in {\rm Harm}_D^4(Y^{11})$ be the orthogonal 
projection onto ${\rm Harm}_D^4(Y^{11})$. 
Then the Hilbert transform 
acts as  $T i^* G_4^N =-i^* \ast G_4^D$, and $Ti^* \ast G_4^D= i^* G_4^N$. 

\noindent $(ii)$ 
$\langle G_4^N~,~ G_4^D \rangle_{L^2} =
|| G_4^N ||_{L^2}~ ||G_4^D||_{L^2} \cos \theta$, where ${\rm cos}^2 \theta$ is the eigenvalue of the 
composition ${\sf proj}_D \circ {\sf proj}_N: {\rm Harm}_D^4(Y^{11}) \to {\rm Harm}_D^4(Y^{11})$. 
\label{prop mix}
\end{proposition}
This is the ``mixing" which complements Proposition \ref{prop pure}. 
We now illustrate with examples in M-theory.

\paragraph{Disk bundles over ten-manifolds.}
Consider the complex six-dimensional projective space $\C P^6$.
 Define a one-parameter 
family of compact Riemannian manifolds with boundary 
$Z^{12}_r:=\C P^6 - \mathbb{B}_r(x)$, where $\mathbb{B}_r(x)$ is an open ball of radius $r \in (0, \pi/2)$ 
centered at a point $x$ in $\C P^6$. 
The nontrivial cohomology groups of 
$Z_r^{12}$ are 
\(
H^{2k}(Z_r^{12};\R) \cong H^{12-2k}(Z_r^{12}, \partial Z_r^{12};\R)=\R; \quad k=0, \cdots, 5.
\)
The twelve-manifold $Z_r^{12}$ is in fact a 2-disk bundle over $\C P^5$, and the boundary $\partial Z_r^{12}$ is 
homeomorphic to the eleven-sphere $S^{11}$.  
For each $1 \leq k \leq 5$, harmonic $2k$-fields satisfying Neumann and Dirichlet 
boundary conditions can be constructed \cite{Sh}.
Let $h_t: S^{11} \to S^{11}_t$ the diffeomorphism of the unit sphere with the hypersurface at 
constant distance $t$ from $\C P^5$. 
Let $H: S^{11} \to \C P^5$ be the Hopf fibration and let $v$ be the vector field on 
$\C P^6$ which restricts on each hypersurface to the pushforward by $h_t$ of the
unit vector field in the Hopf direction on $S^{11}$. Define $\alpha$ to be the dual to $v$, and let 
$\tau=dt$ be the 1-form dual to $\partial/\partial_t$ and define $\eta$ to be the two-form which 
restricts on each $S^{11}_t$ to $(H \circ h_t)^* \eta_{\C P^5}$, where 
$\eta_{\C P^5}$ is the standard symplectic form on $\C P^5$. 
Away from $\C P^5$ the manifold $Z_r^{12}$ is topologically a product $S^{11} \times I$, 
so exterior derivatives can be computed as in $S^{11}$. Thus $d\alpha = - 2\eta$ and 
$\eta$ and $\tau$ are closed. 
Closed and co-closed $4$-forms $G_4^N$ and 
$G_4^D$ satisfying Neumann and Dirichlet conditions, respectively
\(
G_4^N: = f_N(t) \eta^2 + g_N(t) \alpha \wedge \eta \wedge \tau\;,
\qquad \qquad
G_4^D:= f_D (t) \eta^2 + g_D(t) \alpha \wedge \eta \wedge \tau\;,
\)
can be constructed for functions $f$ and $g$.
The angle $\theta$ between $G_4^N$ and $G_4^D$ is in 
Proposition \ref{prop mix} 
with $\theta$ given by \cite{Sh}
$
\cos \theta = (1- \sin^6 r)/[(1+ \sin^6 r)^2 + \frac{1}{2}\sin^6 r]^{1/2}$.
The same result holds for the case when the Euler class of the two-disk bundle 
over $\C P^5$ is varied. The $\mathbb{D}^2$ bundle with Euler class $m$ over
$\C P^5$ has boundary $L(m,1)$, the lens space which is the quotient of 
the sphere $S^{11}$ by the action of $\Z_m$ given by 
$
e^{2\pi i/m} \cdot (z_0, \cdots, z_6)=\left(
e^{2\pi i/m} z_0, \cdots, e^{2\pi i/m} z_6
\right)$.
The result for the angle is the same as above and is independent of the Euler class 
$m$. As $r \to 0$, $\theta \to 0$ so that the two forms are orthogonal in this case.

\subsection{Effect of including the M5-brane}
\label{sec m5}

Consider the M5-brane worldvolume $W^6$ embedded in eleven-dimensional 
spacetime $\iota: W^6 \to Y^{11}$. Take $W^6=S^5 \times I$ and 
$Y^{11}=M^{10} \times I$, where $I$ corresponds to compact time 
direction (we could also use $\R$ in place of the interval $I$). 
We identify a tubular neighborhood of $S^5$ in $M^{10}$ with the total space
of the normal bundle $N \to S^5$. The unit sphere bundle of radius $r$ , 
$X^9=S_r(N)$ is an associated $S^4$ bundle $\pi : X^9 \to S^5$. Now
we can construct in ten dimensions, in analogy to what was done in eleven 
dimensions in \cite{DFM}, a ten-manifold $M^{10}_r$ with boundary 
$X^9$ by removing the disk bundle of radius $r$, $M_r^{10}=M^{10} -\mathbb{D}(N)$.

\vspace{3mm}
\paragraph{Example: Grassmannians.}
Consider the Grassmannian of oriented two-planes in Euclidean space $\R^7$,
${\rm Gr}_2\R^7=SO(7)/SO(5)$, which is a ten-manifold 
with Riemannian submersion metric induced by the 
 bi-invariant 
metrics on the Lie groups $SO(7)$ and $SO(5)$. 
This has a subGrassmannian ${\rm Gr}_1\R^6=SO(6)/SO(5)$, the Grassmannian 
of oriented lines in $\R^6$, which is just the unit five-sphere $S^5$. 
Consider the one-parameter family of manifolds
$
M^{10}_r:= {\rm Gr}_2 \R^{7} -\nu_r ({\rm Gr}_1 \R^6)$,
where $\nu_r({\rm Gr}_1\R^6)$ is the open tubular neighborhood of radius 
$r$ around ${\rm Gr}_1 \R^6$, which is topologically the unit tangent bundle 
of $S^5$.  
$M_r^{10}$ is a $\mathbb{D}^2$ bundle over the eight-manifold 
${\rm Gr}_2 \R^6=SO(6)/SO(4)$.  
The boundary $\partial M_r^{10}$ is homeomorphic
to the unit tangent bundle $US^5$. This has the same rational cohomology as
$S^5 \times S^4$, so all of this cohomology is interior except 
$H^4(M^{10}_r;\R)$ and so $H^4(M_r^{10}, \partial M_r^{10};\R)$ 
could potentially have a 1-dimensional boundary subspace. 
The Poincar\'e duality angle $\theta$ between the concrete realizations of 
$H^4(M_r^{10};\R)$ and $H^4(M_r^{10}, \partial M_r^{10};\R)$ is given by 
\cite{Sh}
$
\cos \theta =(1-\sin^5 r)/[
(1+ \sin^5 r)^2 + \frac{1}{6} \sin^5 r]^{1/2}$.
Again $\theta \to 0$ as $r \to 0$ so that the forms become orthogonal. 


\begin{proposition}
Even on a closed eleven-manfold, 
there is a mixing between Dirichlet and Neumann field strengths $G_4$ (or C-fields) 
due to the presence of the M5-brane worldvolume  
\end{proposition}

We now consider general six- and eleven-manifolds. 
Let $Y^{11}$ be a closed 
smooth oriented Riemannian eleven-manifold and let 
$M^6$ be a closed submanifold of codimension five representing the 
worldvolume of the M5-brane. Define the compact Riemannian 
eleven-manifold $Y^{11}_r:= Y^{11} - \nu_r(M^6)$, where 
$\nu_r(M^6)$ is the open tubular neighborhood of radius $r$ 
about $M^6$. The radius is taken to small enough so that the 
ten-dimensional boundary $\partial Y^{11}_r$ is smooth. 
Let $\theta$ be the Poincar\'e duality angle of $Y^{11}_r$ in 
dimension 4. We are interested in the behavior of $\theta$ as $r \to 0$. 
Examples show that $\theta \sim \mathcal{O}(r^m)$, where 
$m>0$. In fact in \cite{Sh} this is conjectured to hold in general 
with $m$ the codimension of $M$ in $Y$, being five in our case. 
The physical counterpart of this conjecture is then 

\begin{conjecture}
For a general M5-brane worldvolume in a general eleven-dimensional 
manifold, the mixing between 
$H^4(M^{10}_r;\R)$ and so $H^4(M_r^{10}, \partial M_r^{10};\R)$ 
is nonzero in general, and vanishes in the limit when the size of the tubular
is very small. Thus for a macroscopic M5-brane the effect of the Poincar\'e duality angle is visible, 
and the effect disappears in the microscopic limit.  
\end{conjecture}


\section{Global aspects: Bundles and the phase of the partition function}

We have so far considered local questions related to differential forms 
and corresponding differential equations. In this section we shift to more global 
aspects, that is to bundles and to integrals of differential forms. We first consider the $E_8$ bundle 
and then consider the phase of the partition function.

\subsection{$E_8$ gauge theory}
\label{sec e8}
In this section we consider the boundary conditions on the bundles involved, 
especially the $E_8$ bundle on $Y^{11}$, and consider conditions and 
obstructions for extensions.
In \cite{Zu} the case of an abelian gauge theory on 
a 4-dimensional manifold with a boundary was considered. 
We generalize to the case of $E_8$ bundle in eleven dimensions.
Although we discuss mainly the eleven-dimensional case, 
our results apply to other dimensions. 

\paragraph{Extension of $E_8$ bundles.}
Let $(P, A)$
 be a principal bundle $P$ with connection $A \in {\rm Conn}(P)$
  on $Y^{11}$ and 
$(P_\partial, A_\partial)$ be
a principal bundle on the boundary $\partial Y^{11}$ with the same structure
group. 
Every bundle $P$ on $Y^{11}$ yields by pullback a bundle 
$P_\partial$ on $\partial Y^{11}$, i.e. $P_{\partial}=i^* P$. 
However the converse is not always true; 
not every  bundle $P_\partial$ on the boundary always the pullback of some  
bundle $P$ on $Y^{11}$. We consider the spaces ${\rm Prin}(Y^{11})$,
${\rm Prin}(\partial Y^{11})$, ${\rm Prin}(Y^{11}, \partial Y^{11})$
 of 
principal $E_8$ bundles over $Y^{11}$, over $\partial Y^{11}$, 
and those over $Y^{11}$ which vanish on the boundary $\partial Y^{11}$, respectively.
Since $BE_8 \sim K(\Z,4)$ in our range of dimensions
we have 
 isomorphisms ${\rm Prin}(Y^{11}) \buildrel{c}\over{\cong} H^4(Y^{11};\Z)$ 
 and ${\rm Prin}(\partial Y^{11}) \buildrel{c_\partial}\over{\cong} H^4(\partial Y^{11};\Z)$.
 Furthermore, we have the following commutative
diagram
\(
\xymatrix{
H^3(\partial Y^{11};\Z) 
\ar[r]^{\hspace{-5mm} \delta}
&
H^4(Y^{11}, \partial Y^{11};\Z)
\ar[r]
&
H^4(Y^{11};\Z)
\ar[r]^{\iota^*}
&
H^4(\partial Y^{11};\Z)
\ar[r]^{\hspace{-5mm} \delta}
&
H^5(Y^{11},\partial Y^{11};\Z)
\\
&
{\rm Prin}(Y^{11}, \partial Y^{11})
\ar[r]
\ar[u]^c
&
{\rm Prin}(Y^{11})
\ar[r]^{\iota^*}
\ar[u]^c
&
{\rm Prin}(\partial Y^{11})
\ar[u]^{c_\partial}
&
}.
\label{Prin diag}
\)
The horizontal arrows are exact and the vertical maps are isomorphisms.
The first map in the second line associates to every relative bundle 
the underlying bundle, and the second arrow associates to every
bundle $P$ its pullback bundle $P_\partial = \iota^*P$. 
By exactness in the diagram, 
a boundary bundle $P_\partial \in {\rm Prin}(\partial Y^{11})$ is the 
pullback of a bundle $P \in {\rm Prin}(Y^{11})$ if and only if 
\(
\delta (c_\partial (P_\partial))=0\;,
\)
so that the obstruction to the extendibility of $P_\partial$ is a class 
of $H^5(Y^{11}, \partial Y^{11};\Z)$.
When this is satisfied, $P_\partial$ is in general the pullback of more than 
one bundle $P$ on $Y^{11}$, that is $P$ has several extensions to $Y^{11}$. 
The extensions are parametrized by the group of relative bundles in a fashion which
is one-to-one provided $H^3(\partial Y^{11};\Z)=0$. 

\paragraph{Extensions of gauge transformations.}
Let $\cG$ denote the group of gauge transformations 
\footnote{i.e. gauge group for 
a mathematician.} 
of the $E_8$ bundle. 
Starting with a gauge transformation $U \in \cG (\partial Y^{11})$
we get by pullback a gauge transformation
$U_\partial \in \cG(\partial Y^{11})$, that is, $U_\partial =i^* U$.
The converse is not true, that is not every boundary gauge transformation 
$U_\partial$ the pullback of some gauge transformation 
$U$. In \cite{Zu} the obstruction to such an extension was identified for the 
abelian case in four dimensions, and so here we work out the analog for 
the $E_8$ gauge theory in eleven dimensions (following similar arguments). 
Consider the gauge transformation class groups
$$
{\cC}(Y^{11})=\cG (Y^{11})/\cG_c(Y^{11})\;,
\quad 
{\cC}(\partial Y^{11})=\cG (\partial Y^{11})/\cG_c(\partial Y^{11})\;,
\quad 
{\cC}(Y^{11}, \partial Y^{11})=
\cG (Y^{11}, \partial Y^{11})/\cG_c(Y^{11}, \partial Y^{11})\;,
$$
where the factor groups are connected components to the corresponding identity 
gauge transformations.
Now we have an analog of the diagram
\eqref{Prin diag}, 
\(
\xymatrix{
H^2(\partial Y^{11};\Z) 
\ar[r]^{\hspace{-5mm} \delta}
&
H^3(Y^{11}, \partial Y^{11};\Z)
\ar[r]
&
H^3(Y^{11};\Z)
\ar[r]^{\iota^*}
&
H^3(\partial Y^{11};\Z)
\ar[r]^{\hspace{-5mm} \delta}
&
H^4(Y^{11},\partial Y^{11};\Z)
\\
&
{\cC}(Y^{11}, \partial Y^{11})
\ar[r]
\ar[u]^q
&
{\cC}(Y^{11})
\ar[r]^{\iota^*}
\ar[u]^q
&
{\cC}(\partial Y^{11})
\ar[u]^{q_\partial}
&
}\;.
\)
Here $q$ assigns to each gauge transformation $U$ its characteristic class
$q(U)$.
By exactness, a gauge transformation class $[U_\partial]\in \cC(\partial Y^{11})$
is the pullback of a gauge transformation class $[U]\in \cC(Y^{11})$ 
if an only if 
\(
\delta (q_\partial ([U_\partial]))=0\;.
\label{gauge cond}
\)
Hence the obstruction to the extendibility of $[U_\partial]$ is a class of
$H^4(Y^{11}, \partial Y^{11};\Z)$. 
When $[U_\partial]$ satisfies \eqref{gauge cond}, $[U_\partial]$
is the pullback of possibly several gauge transformation classes
$[U]\in \cC(Y^{11})$, that is $[U_\partial]$ has possibly several extensions
to $Y^{11}$. These extensions are parametrized by the group of relative
gauge transformations $\cG(Y^{11},\partial Y^{11})$, which is 
a one-to-one correspondence when $H^2(\partial Y^{11};\Z)=0$.

\vspace{3mm}
The boundary condition on these bundles is
$\i^* P= P_\partial$.  Also, a natural choice for the connection is 
\(
i^*A=A_\partial\;,
\label{conn BC}
\)
 where $A_\partial \in {\rm Conn}(P_\partial)$
is a fixed connection.
These boundary conditions are not preserved under the action of 
gauge transformations $\cG(Y^{11})$ on $Y^{11}$. However, they are 
preserved by the group of relative gauge transformations
$\cG(Y^{11}, \partial Y^{11})$. 
The allowed variations on the connection $A \in {\rm Conn}(P)$ 
preserving the boundary condition \eqref{conn BC}
are Lie algebra-valued 1-forms $\delta_g A \in \Omega^1(Y^{11})\otimes \frak{e}_8$
such that $\iota^*\delta_g A=0$, that is 
$\delta_g A\in \Omega^1_D(Y^{11})\otimes \frak{e}_8$ satisfies the Dirichlet 
boundary conditions. 
Our previous constructions can be extended to the nonabelian case. 
Define $d_\partial^A=d_\partial + A_\partial$, and take $d^{{}^*A}_\partial$ be 
its adjoint. For instance, the {\it nonabelian Hilbert transform} will now become
$T_A:= d_\partial^A \Lambda^{-1}= T + A_\partial \Lambda^{-1}$, and 
similarly for other entities. 

\vspace{3mm}
We summarize our findings in this section in

\begin{theorem}
$(i)$ An $E_8$ bundle $P_\partial \in {\rm Prin}(\partial Y^{11})$ is the 
pullback of a bundle $P \in {\rm Prin}(Y^{11})$ if and only if 
$
\delta (c_\partial (P_\partial))=0$, 
so that the obstruction to the extendibility of $P_\partial$ is a class 
of $H^5(Y^{11}, \partial Y^{11};\Z)$.
When this is satisfied,
the extensions are parametrized by the group of relative bundles in a fashion which
is one-to-one provided $H^3(\partial Y^{11};\Z)=0$. 

\noindent $(ii)$ A gauge transformation class $[U_\partial]\in \cC(\partial Y^{11})$
is the pullback of a gauge transformation class $[U]\in \cC(Y^{11})$ 
if an only if 
$\delta (q_\partial ([U_\partial]))=0$, so
that the obstruction to the extendibility of $[U_\partial]$ is a class of
$H^4(Y^{11}, \partial Y^{11};\Z)$. 
When this is satisfied, the extensions are parametrized by the group of relative
gauge transformations $\cG(Y^{11},\partial Y^{11})$, which is 
a one-to-one correspondence when $H^2(\partial Y^{11};\Z)=0$. 
\end{theorem}

\subsection{The phase of the partition function via the adiabatic limit}
\label{sec eta}
In this section we consider the effect of having the boundary $\partial Y^{11}$ on the 
phase of the partition function, given by the exponentiated eta-invariants \cite{DMW}
(cf. equation \eqref{exp ac})
\(
\Phi(C_3)= \exp \left[ 2\pi i \left(\frac{1}{2} \overline{\eta}_{E_8} + \frac{1}{4} \overline{\eta}_{RS}
\right)\right]\;,
\label{eq phase}
\)
where $\overline{\eta}:=\frac{1}{2}(\eta + h)$, each for an $E_8$ bundle and the 
Rarita-Schwinger bundle, and $h$ the number of zero modes of the corresponding 
twisted Dirac operator.
In \cite{DMW-sig} this is written in terms of the eta
invariant corresponding to the signature operator on $Y^{11}$. Thus, the phase can be studied
either using Dirac operators or the signature operator (the latter can in a sense also be 
considered a Dirac operator).  The phase \eqref{eq phase}
is the result of the use of the APS index theorem for a twelve-manifold $Z^{12}$ with 
a boundary $\partial Z^{12}=Y^{11}$ on the phase in twelve dimensions written in 
terms the Atiyah-Singer index \cite{Flux}. The dimensional reduction to ten-dimensions is 
performed in \cite{MS} \cite{S-gerbe} \cite{DMW-sig} via the adiabatic limit of the eta
invariants which result in expressions for the phase in ten dimensions.

\vspace{3mm}
We would like to consider the case when $Y^{11}$ has a non-empty boundary 
$\partial Y^{11}$. So we no longer assume that $Y^{11}$ is itself a boundary, 
and the phase \eqref{eq phase} is taken as a starting point, as in \cite{DFM}.
In M-theory we are in practice interested in eleven-manifolds which are 
decomposable into a product or which are total spaces of bundles. Therefore, 
we consider eleven-manifolds with boundary which are products or fiber bundles.
Note that the product of a manifold-with-boundary with a manifold without boundary 
is a manifold with boundary, and similarly for bundles. Therefore, we will consider 
bundles with total space $Y^{11}$ where either the fiber or the base has a boundary, 
which makes $Y^{11}$ itself a manifold with boundary.

\paragraph{I. The base space with a boundary.}
Consider a bundle $N^n \hookrightarrow Y^{11} \buildrel{\pi}\over{\longrightarrow}{X^{11-n}}$, 
where the fiber $N^n$ is a closed compact Spin $n$-manifold with metric $g_N$, 
and the base 
$X^{11-n}$ is a compact Spin manifold with boundary  $\partial X^{11-n}$
and metric $g_X$. 
This makes the total 
space $Y^{11}$ into  a manifold with boundary on which we take a family of 
submersion metrics $g_\epsilon$ of the form $g_\epsilon= g_N + \frac{1}{\epsilon^2} \pi^*g_X$
as $\epsilon \to 0$. 
The Spin bundle of the total space is given in terms of the Spin 
bundles of the base and the fiber as $S(Y^{11})=\pi^* S(X^{11-n}) \otimes S(N^n)$. 
We have a  total Dirac operator $D_\epsilon^Y$ defined on $Y^{11}$, a 
 boundary Dirac operator $D_\epsilon^{\partial Y}$ on $\partial Y^{11}$,
 and a family of Dirac operators $D^N$ along the fibers. 
Assuming that the Dirac operators along the fibers $N^n$ 
are invertible gives that the Dirac operator along the 
boundary $\partial Y^{11}$ is invertible for small $\epsilon$. 
This makes the APS problem for the Dirac operator on $Y^{11}$ self-adjoint and so 
there is an eta-invariant. 
Let $\eta(D_\epsilon^Y)$ denote the eta-invariant of $D_\epsilon^Y$ with the APS
boundary conditions. Then 
the adiabatic limit in this case $\lim_{\epsilon \to 0} \overline{\eta} (D_\epsilon^Y)=
\lim_{\epsilon \to 0}\frac{1}{2}\eta (D_\epsilon)$ exists as a real number and,
from \cite{D},  
is given by 
\(
\lim_{\epsilon \to 0} \overline{\eta}(D_\epsilon)=
\int_{X^{11-n}} \widehat{A}(R^X) \wedge \widehat{\eta}\;,
\)
where $R^X$ is the curvature of $g_X$ and $\widehat{\eta}$ is the (normalized)
Bismut-Cheeger
$\eta$-form. 
The formula is exactly the same as the case with no boundary \cite{BC1} -- applied to 
M-theory in \cite{MS} and \cite{S-gerbe} --
so that the boundary does not contribute in this case. 
The above limit can also be taken in the presence of an $E_8$ vector bundle. 
In this case we have an analogous conclusion to that of \cite{DFM}:

\begin{proposition}
The dimensional reduction of the phase of the partition function of M-theory on a manifold with 
boundary is not changed when the base manifold has a boundary. That is , the phase of the 
partition function in this case is not different from the case when $\partial Y^{11}=\emptyset$. 
\end{proposition}

\paragraph{II. The fiber with a  boundary.}

Now consider a bundle 
$N^{n} \to Y^{11} \buildrel{\pi}\over{\longrightarrow} X^{11-n}$
where the fiber $N^{n}$ is a compact manifold with boundary 
$\partial N^{n}$. Assume that the vertical tangent bundle 
$TN^{n} \subset TY^{n}$ is equipped with a Spin structure and a
Riemannian metric, such that for each fiber $N^{n}$
the induced metric $g^N$ splits isometrically as 
$dz^2 + g^{\partial N}$ on $[0,1]\times \partial N^{n}\subset N^{n}$. 
Now consider an $E_8$ vector bundle 
$E \to Y^{11}$ with connection $\nabla^E$ such that the 
restriction to $[0,1]\times \partial Y^{11}$ is the pullback of some connection
on $E_{\partial Y}$ via the natural projection. Two cases are considered
depending on the parity of the dimension of the fiber.

 \paragraph{1. Fiber is odd-dimensional.} 
 Consider $Y^{11}$ to be the total space of the bundle 
 $N^{2l+1} \to Y^{11} \buildrel{\pi}\over{\longrightarrow} X^{10-2l}
$ with odd-dimensional fiber. 
 $Y^{11}$ equipped with an exact b-metric
  ${}^bg_Y$ (see \cite{Mel} for how such metrics are defined).
Let $x \in C^\infty (M^{10})$ be a distinguished
defining function for the boundary $M^{10}=\partial Y^{11}$
so that the metric is of the form 
$
{}^bg_Y= \left( \frac{dx}{x}\right)^2 + g_M$.
The corresponding contangent bundle 
${}^bT^*Y^{11}$ decomposes orthogonally
over the boundary
$
{}^bT^*_{\partial Y}Y^{11}=\R\left(\frac{dx}{x} \right)\oplus T^* \partial Y^{11}
$.
Consider a family $\{D_x\}_{x \in X}$ of Dirac operators 
on the fibers $N_x$, and let $\{D_0\}$ be the boundary family. 
From the work of Atiyah-Singer this
 defines an index class in the K-theory of the base ${\rm Ind}(D)\in K^1(X)$. 
 A formula for the corresponding Chern character ${\rm ch}({\rm Ind}(D))\in H^{\rm odd}(X)$ 
 was conjectured in \cite{BC2} under the assumption that all the operators induced on 
 the fiber boundary $\partial N$ are invertible, and under the APS boundary conditions.
 This is proved in \cite{MP} with the invertibility assumption 
 relaxed. 
Let $D$ be the family of Dirac operators on the fiber and let $D_0$ be the boundary family
(that is on $\partial N^{2l+1}$). The index class ${\rm Index} (D) \in K^1(X)$ has Chern 
character \cite{MP}
\(
{\rm ch}({\rm Ind}(D, {\sf P}))=
\int_{Y/X} \widehat{A}(Y/X) {\rm ch}(E) 
-\frac{1}{2}\widehat{\eta}_{{\rm odd}, {\sf P}} ~\in H^{\rm odd}(X)
\label{eq mp1}
\)
where ${\rm ch}(E)$ is the Chern character of the twisting curvature of the vector bundle 
and ${\sf P}$ is the spectral section projection.  

\paragraph{2. Fiber is even-dimensional.} 
Consider the bundle  $N^{2l} \to Y^{11} \buildrel{\pi}\over{\longrightarrow} X^{11-2l}$,
where the fiber is even dimensional with a boundary, $\partial N^{2l}$. 
As above we assume that the tangent bundle to the fibers $TN^{2l} \subset TY^{11}$
is equipped with a Spin structure. 
We take the  Riemannian metric $g^N$ to be a product 
$g^N=dz^2 + g^{\partial N}$ on $[0,1] \times \partial N^{2l} \subset N^{2l}$. 
Consider the Spin bundle along the fiber $S_N=S_N^+ \oplus S_N^-$,
where $S_N^\pm$ are positive and negative chirality parts. 
Again take $E\to Y^{11}$ to be an $E_8$ bundle over $Y^{11}$ with a connection
which restricts on $[0,1] \times \partial Y^{11}$ to the pullback of some connection 
on $E_{\partial Y^{11}}$ via the natural projection (see section \ref{sec e8}). 
Consider the twisted Dirac operator along the fiber 
$
D_N^+: S_N^+ \otimes E \longrightarrow S_N^- \otimes E
$
with the Atiyah-Patodi-Singer boundary conditions. 
Then, from \cite{BC1} \cite{BC2}, 
the family $D_N^+$ determines a continuous family of Fredholm operators
with index ${\rm Ind} D_N^+$ an element in the K-theory of the base $K^0(X^{11-2l})$ with 
Chern character 
\(
{\rm ch}({\rm Ind}(D_N^+))=\int_{N^{2l}} \widehat{A} (R^N) {\rm ch}(E) - 
\widehat{\eta}\;,
\label{eq mp2}
\)
where $R^N$ is the curvature of the connection on $TN^{2l}$. 

\begin{proposition}
Consider the partition function for M-theory on a manifold with boundary
which of the form of a bundle with base space a closed manifold and a
fiber and odd-, respectively, even-dimensional manifold with boundary. 
Then  
the phase of the partition function can be reduced to ten dimension via the adiabatic
limit to the (logarithm of the) formulas \eqref{eq mp1} and \eqref{eq mp2},
respectively, provided we impose the appropriate boundary conditions as above. 
\end{proposition}

 
 \subsection{The new Chern-Simons term from the geometric correction to the index}
 \label{sec cs}
 We have discussed the importance of the boundary conditions for fields, bundles
 and for applying the index theorem for Dirac operators on an eleven-dimensional
 manifold with boundary.  In this section we consider the effect of considering 
 boundary conditions which are not of APS type, that is the manifold is not
 of the form of a cylinder near the boundary. 
 In a previous paper we have shown that the topological action, and hence the 
 phase of the partition function, can be recast in terms of the signature operator
 in place of the Dirac operator \cite{DMW-sig}.
 Writing the action in terms of the signature allows us to 
 make use of the results of Gilkey \cite{G} to find geometric corrections to the 
 index formula. 
 In this case, the APS index formula can be written as 
 \(
 {\rm Index}(D, Y^{11})= {\rm Index}(D, Z^{12}) + \overline{\eta}(Y^{11}) +  S[Y^{11}]\;, 
  \)
 where the first and second terms on the right hand side are the usual terms in the 
 APS index formula, that is the index for the case when there is no boundary and the 
 correction from the eta-invariant and zero modes. The third term is an integral over 
 $Y^{11}$ of some Chern-Simons term constructed in \cite{G} and used 
 for the signature operator in 
 \cite{DMW-sig} for the case when $Z^{12}$ is a disk bundle.  
 This term arises when the metric near the boundary is not a product metric
 and can be calculated from the connection determined by a suitable choice of the 
 normal to the boundary. The connection $\omega_Y$ decomposes 
 into tangential and normal components $\omega^T_Y$ and $\omega^N_Y$, respectively,
 the latter being the second fundamental form.  
This is covariant under transformations of frames. The difference between characteristic 
polynomials $P(\Omega_Y)$ and $P(\Omega_Y^T)$ corresponding to the curvatures 
$\Omega_Y$ and $\Omega_Y^T$
of the
connections $\omega_Y$ and $\omega_Y^T$, respectively, is an exact form 
$dQ( \Omega_Y, \Omega_Y^T) =P(\Omega_Y) - P(\Omega_Y^T)$, where 
$Q(\Omega_Y, \Omega_Y^T)$ is a Chern-Simons form. The surface term is then 
\cite{G}
$
S[Y^{11}]= -  \int_{Y^{11}} Q((\Omega_Y, \Omega_Y^T))$.
Explicitly, the eleven-dimensional gravitational Chern-Simons term is
\(
L_{CS}=k \int_{Y^{11}} CS_{11}(\omega_Y)=6 k \int_{Y^{11}}  \int_0^1 dt {\rm Tr} 
\left[\omega_Y \wedge(td\omega_Y + t^2 \omega_Y \wedge\omega_Y)
\wedge \cdots \wedge (td\omega_Y + t^2 \omega_Y \wedge\omega_Y)  \right]\;,
\)
where $k$ is an integer, a condition from the requirement of independence of the 
twelve-dimensional extension. Note that the main ingredient in Horava's 
proposal \cite{Ho} 
for a holographic field theory describing M-theory is such a gravitational 
Chern-Simons action. Thus, we see that this action is part of the action 
for M-theory that we get from the C-field. 
We emphasize that we get this gravitational Chern-Simons term in 
addition to the other terms that we already have in the (exponentiated) 
action had we not considered the geometric correction to the index formula. 
We therefore have 

\begin{proposition}
$(i)$ The phase of the M-theory partition function in general boundary conditions
leads to gravitational Chern-Simons eleven-form $CS_{11}$ as correction.

\noindent $(ii)$ The holographic Chern-Simons description of M-theory is a phase in 
the index-theoretic approach. 
\end{proposition}

\paragraph{Remarks on the case when $Y^{11}$ has a boundary.}
In contrast to 2+1 dimensions, the higher dimensional Chern-Simons theory
does have local, physical degrees of freedom \cite{BGH}.
Again, in the presence of a boundary we
 take the Chern-Simons term in eleven dimensions as a starting 
point and ask to which quantity it reduces on the boundary. 
As pointed out in \cite{Ho} these correspond to edge states, given by the 
$E_8$ super Yang-Mills, on the 
Horava-Witten boundary and are analogous to similar states in Chern-Simons 
gauge theory (see for instance  \cite{BCE} and references therein). We make a 
few remarks for future investigation.

\vspace{2mm}
\noindent {\it 1.} It would be interesting to see if the Chern-Simons term we found comes
with a fermionic term which makes it supersymmetric. We expect this to be
the case as eleven-dimensional 
supergravity can be written in terms of supersymmetric Chern-Simons
theory (see \cite{TZ}). 

\vspace{2mm}
\noindent {\it 2.} The geometric correction to the index leading to the Chern-Simons term 
comes from 
the second fundamental 
form \cite{G}. We expect this to be explain 
 the surface term proposed in \cite{Mo1} to modify the Horava-Witten set-up
\cite{HW}. 
For the spinors, the obvious boundary condition used in \cite{HW} has to be 
modified to a more elaborate boundary condition, involving projection operators
which depend on the gaugino expectation value \cite{Mo1}.

\vspace{2mm}
\noindent {\it 3.} The theory to which the Chern-Simons theory restricts 
on the boundary seems to be a higher-dimensional generalization 
of the WZW model (see \cite{BGH} \cite{LMNS}). We 
plan to investigate this elsewhere, building on \cite{S-target}.


\end{document}